\documentclass[twocolumn,aps,prb,floats,floatfix,subscriptaddress,showpacs]{revtex4}

\usepackage{graphicx}
\begin{document}
%
%Listing the affiliations first sets the order - allows us to put CIAR last
%
\affiliation{Condensed Matter Sciences Division, Oak Ridge
National Laboratory, Oak Ridge, TN 37831}

\affiliation{Hahn-Meitner Institut, Glienicker Str. 18, Berlin
D-14607, Germany}

\affiliation{Department of Physics and Astronomy, University
College London, UK}

\affiliation{Department of Physics, University of Virginia,
Charlottesville, Virginia 22904, USA}

%
%Now come the authors with their affiliation
%
\author{M.D. Lumsden}
\affiliation{Condensed Matter Sciences Division, Oak Ridge
National Laboratory, Oak Ridge, TN 37831}
\author{S.E. Nagler}
\affiliation{Condensed Matter Sciences Division, Oak Ridge
National Laboratory, Oak Ridge, TN 37831}
\author{B.C. Sales}
\affiliation{Condensed Matter Sciences Division, Oak Ridge
National Laboratory, Oak Ridge, TN 37831}
\author{D.A. Tennant}
\affiliation{Hahn-Meitner Institut, Glienicker Str. 100, Berlin
D-14109, Germany}

\affiliation{ISIS Facility, Rutherford Appleton Laboratory,
Chilton, Didcot OX11 0QX, U.K.}

\affiliation{Clarendon Laboratory, University of Oxford, Parks
Road, Oxford, OX1 3PU, U.K.}
\author{D.F. McMorrow}

\affiliation{London Centre for Nanotechnology and Department of
Physics and Astronomy, University College London, London WC1E 6BT,
U.K.}

\affiliation{Materials Research Department, Ris\o\ National
Laboratory, 4000 Roskilde, Denmark}

\affiliation{ISIS Facility, Rutherford Appleton Laboratory,
Chilton, Didcot OX11 0QX, U.K.}

\author{S.-H. Lee}
\affiliation{Department of Physics, University of Virginia,
Charlottesville, Virginia 22904, USA}

\affiliation{NIST Center for Neutron Research, National Institute
of Standards and Technology, Gaithersburg, MD 20899}
\author{S. Park}
\affiliation{HANARO Center, Korea Atomic Energy Research
Institute, Daejeon, Korea }

\affiliation{NIST Center for Neutron Research, National Institute
of Standards and Technology, Gaithersburg, MD 20899}
\title{Magnetic Excitation Spectrum of the Square Lattice S=1/2 Heisenberg Antiferromagnet K$_2$V$_3$O$_8$}
\date{\today }
\begin{abstract}
We have explored the magnetic excitation spectrum of the S=1/2
square lattice Heisenberg antiferromagnet, K$_2$V$_3$O$_8$ using
both triple-axis and time-of-flight inelastic neutron scattering.
The long-wavelength spin waves are consistent with the previously
determined Hamiltonian for this material.  A small energy gap of
72$\pm$9 $\mu$eV is observed at the antiferromagnetic zone center
and the near-neighbor exchange constant is determined to be
1.08$\pm$0.03 meV. A finite ferromagnetic interplanar coupling is
observed along the crystallographic c-axis with a magnitude of
J$_c$=-0.0036$\pm$0.0006 meV.  However, upon approaching the zone
boundary, the observed excitation spectrum deviates significantly
from the expectation of linear spin wave theory resulting in split
modes at the ($\pi$/2,$\pi$/2) zone boundary point. The effects of
magnon-phonon interaction, orbital degrees of freedom, multimagnon
scattering, and dilution/site randomness are considered in the
context of the mode splitting. Unfortunately, no fully
satisfactory explanation of this phenomenon is found and further
theoretical and experimental work is needed.

\end{abstract}
\maketitle

\section{Introduction}

Quantum magnetism has been a topic of considerable interest for
many decades \cite{Manousakis}, with particular interest in
two-dimensional systems stimulated by the discovery of high-T$_c$
superconductivity in oxides comprised of CuO$_2$ layers
\cite{highTc}.  Shortly after this discovery, it was realized that
the parent compounds of these superconductors are well described
by the quantum (S=1/2) square lattice Heisenberg antiferromagnet
(QSLHAF) \cite{Anderson,Endoh} and since then this model system
has been the topic of considerable theoretical and experimental
investigation.

%The field of quantum magnetism has been a topic of considerable
%interest for many decades \cite{Manousakis}. However, the field
%experienced a resurgence following the discovery of high-T$_c$
%superconductivity in oxides comprised of CuO$_2$ layers
%\cite{highTc}. Shortly after this discovery, it was realized that
%the parent compounds of these superconductors are well described
%by the quantum (S=1/2) square lattice Heisenberg antiferromagnet
%(QSLHAF) \cite{Anderson,Endoh} and this model system has been the
%topic of considerable theoretical and experimental investigation.

Theoretically, there is growing consensus that the ground state of
the QSLHAF is long-range ordered only at zero temperature. The
dynamics of the QSLHAF are well described by classical linear
spin-wave theory with quantum corrections, in the form of higher
order expansions in 1/S \cite{Igarashi1,Singh}, necessary to
extract accurate physical parameters.  For instance, classical
spin-wave theory accurately reproduces the dynamical structure
factor, S(\textbf{Q},$\omega$), with physical parameters, such as
the spin-wave velocity, spin stiffness, and susceptibility
strongly renormalized by these quantum corrections. Qualitative
deviations from spin-wave theory may be manifest near the
antiferromagnetic zone boundary, as shown from both quantum
Monte-Carlo \cite{Sandvik} and series expansion \cite{Singh,Zheng}
studies.
%The only deviations from
%spin-wave theory occur near the antiferromagnetic zone boundary,
%as shown from both quantum monte carlo \cite{Sandvik} and series
%expansion \cite{Singh,Zheng} studies.
These predictions suggest a
7-9\% dispersion in the magnetic excitations between the 2d zone
boundary points ($\pi$/2,$\pi$/2) and ($\pi$,0). Linear spin-wave
theory predicts no dispersion between these points although recent
extensions of the spin-wave calculation including up to second
order corrections in 1/S show a 2\% dispersion \cite{Igarashi2},
much smaller than that predicted by other theoretical techniques.
The effect of temperature on this model system is to destroy the
T=0 long-range order and the temperature dependence of the
correlation length, which provides a measure of the degree of
order, has been well studied both theoretically
\cite{Chakravarty,Kim,Beard,Elstner}and experimentally
\cite{Yamada,Birgeneau,Greven,Ronnow,Carretta}. The temperature
dependence of the dynamical structure factor has also been the
topic of considerable theoretical effort
\cite{Tyc,Auerbach,Takahashi,Sokol,Sherman,Winterfeldt,Nagao,Wang,Makivic}
with much less experimental effort \cite{cftd}. In recent years
theoretical investigations have been extended to include effects
of an external applied magnetic field
\cite{Yang,Zhitomirsky1,Zhitomirsky2,Syljuasen}. Perhaps the most
dramatic prediction of these studies is a decay of the single
magnon spectrum into a two-magnon continuum for fields
sufficiently close to the saturation field \cite{Zhitomirsky2}.

As is evident in the preceding paragraph, many of the interesting
theoretical studies of the QSLHAF have involved detailed
calculations of the dynamical properties.  These properties are
directly probed by inelastic neutron scattering experiments which
measure the dynamical structure factor, S(\textbf{Q},$\omega$).
Experimental measurements have focussed on the cuprates and
inelastic neutron scattering measurements have been hindered by
the large characteristic energy scale (\emph{J} $\sim$ 1500 K) of
these materials.  Early reactor-based measurements on the
high-T$_c$ parent compound, La$_2$CuO$_4$, were only able to
measure the spin-wave velocity in the long-wavelength limit
\cite{Shirane,Endoh,Aeppli}.
%Early measurements on La$_2$CuO$_4$ using thermal neutrons were unable
%to clearly measure the spin-wave dispersion due to the large
%energy scale but placed a lower bound on the spin wave velocity of
%400 meV-\AA \cite{Shirane,Endoh}. Later measurements using
%neutrons moderated by a hot source were able to clearly measure
%the spin wave velocity to be 850 meV-\AA \cite{Aeppli}.
The abundance of epithermal neutrons at spallation neutron sources
have allowed for measurements of the full dispersion
\cite{Hayden}, resulting in a near-neighbor coupling constant of
136 meV for La$_2$CuO$_4$. Advancement in time-of-flight
instrumentation and visualization software allowed much more
detailed measurements to be performed \cite{Coldea}.  In these
experiments, the zone boundary dispersion was measured and a
strong dispersion was observed between the ($\pi$/2,$\pi$/2) and
($\pi$,0) zone boundary point but in the opposite trend to that
predicted theoretically \cite{Sandvik,Singh,Zheng}.  This trend
was attributed to the presence of a higher order ring exchange
interaction around Cu$_4$O$_4$ square plaquettes \cite{Coldea}, an
interpretation which has been the topic of considerable interest
both theoretically \cite{Peres,Singh2,Katanin} and experimentally
\cite{Toader}. In addition, several other related cuprate model
systems have been studied. One of the best realizations of the
QSLHAF is Sr$_2$CuO$_2$Cl$_2$ where neutron scattering experiments
have produced detailed measurements of the temperature dependent
correlation length \cite{Greven} but no studies of the dynamics
have been performed to date. Another related compound,
Sr$_2$Cu$_3$O$_4$Cl$_2$ is complicated by the presence of two
interpenetrating square Cu$^{2+}$ sublattices with very different
coupling constants (\emph{J}$_I$ $\sim$ 130 meV and
\emph{J}$_{II}$ $\sim$ 10 meV) and frustrated interactions between
the sublattices. \cite{Kim2}. The magnetic excitation spectrum was
measured in detail, for the specific range of temperatures where
this material behaves as a QSLHAF with \emph{J} $\sim$ 10 meV, and
these measurements provided the first experimental evidence for
the theoretically predicted dispersion between the
($\pi$/2,$\pi$/2) and ($\pi$,0) zone boundary points \cite{Kim2}.

Temperature dependent measurements were limited by the large
coupling constants in the cuprates to temperatures considerably
less than \emph{J}. These measurements were extended following the
discovery of the metallo-organic Cu(DCOO)$_2\cdot$4D$_2$O (CFTD)
which has a much smaller coupling constant, \emph{J} $\sim$ 6.3
meV \cite{cftd}. Measurements of the correlation length were
extended to temperatures comparable to \emph{J}
\cite{Ronnow,Carretta}and were found to be in good agreement with
theoretical predictions. Inelastic neutron scattering measurements
on this material \cite{cftd} confirmed the theoretically predicted
dispersion along the zone-boundary \cite{Sandvik,Singh,Zheng} and
also measured the temperature dependence of the dynamical
structure factor. Excitations were found to persist to
temperatures as high as \emph{J}/2 but were found to broaden and
soften upon warming \cite{cftd}.

Experimental studies of the effects of an external magnetic field
have been impossible due to the magnitude of the coupling
constants in known model materials. Interesting theoretical
predictions in the presence of an applied magnetic field require
field strengths near saturation. Using the value of \emph{J}, we
can estimate the field required for saturation for the model
systems mentioned above. The resulting saturation field for
La$_2$CuO$_4$ is $\sim$3800 T, Sr$_2$Cu$_3$O$_4$Cl$_2$ is
$\sim$350 T ($\sim$4500 T) assuming the lower (higher) of the two
coupling constants, and CFTD is $\sim$220 T. Clearly all of these
materials are unsuitable for magnetic field measurements.
Recently, a family of metallo-organics have been synthesized with
much smaller coupling constants giving rise to saturation fields
ranging from 2.25-25 T \cite{Woodward}. Unfortunately, these
materials have rather strong three-dimensional interactions and
are also problematic for inelastic neutron scattering experiments
as it has proven very difficult to grow large, deuterated single
crystals. One particularly attractive material, and the topic of
this communication, is K$_2$V$_3$O$_8$ which has a saturation
field of $\sim$38 T, experimentally achievable for some
experiments.
% Using J=1.08, we get 1.08*2/0.057884=37.5T
% It's unclear if we should use 1.08meV or 1.18*1.08=1.275meV.
% As the field will suppress the quantum fluctuations, it's reasonable
% to expect the 37.5T value.

In this manuscript, we present detailed inelastic neutron
scattering measurements and analysis of the zero-field magnetic
excitation spectrum of K$_2$V$_3$O$_8$.  Such studies are not only
essential to understand future results obtained in the presence of
a finite applied magnetic field but are quite interesting in their
own right. Comparison of the measured spectrum to the predictions
of classical linear spin-wave theory has proven very interesting
in other model systems as emphasized by the opposing trend in the
zone boundary dispersions observed in La$_2$CuO$_4$ \cite{Coldea}
and CFTD \cite{Ronnow}. The zone boundary dispersion observed
K$_2$V$_3$O$_8$ is different than that observed in any of the
previously explored model systems or predicted in theoretical
studies, displaying two modes in the excitation spectrum in the
immediate vicinity of the zone boundary. The remainder of the
paper will be organized in the following manner; we begin by
discussing the properties of K$_2$V$_3$O$_8$ (Section II) followed
by a description of the details of the inelastic neutron
scattering experiments (Section III). The experimental results are
described in Section IV for spin-waves in the long-wavelength
limit which is fully consistent with the prediction of linear
spin-wave theory. In Section V the experimental results near the
zone boundary are described showing the evidence for the two modes
in the excitation spectrum. Several possible sources of the
observed zone boundary mode splitting will be discussed in Section
VI followed by conclusions in Section VII.

\begin{figure}
\centering
\includegraphics[width=\columnwidth,clip]{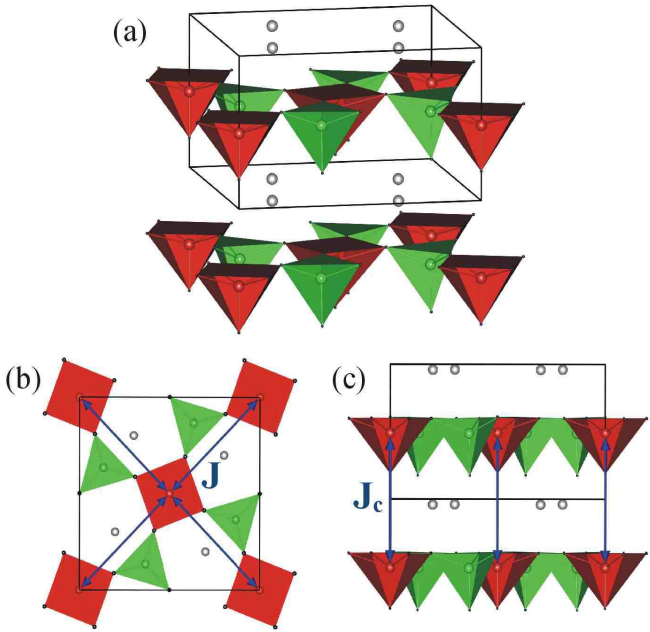}
\caption{(Color online) (a) Crystal structure of K$_2$V$_3$O$_8$
showing red V$^{4+}$O$_5$ pyramids linked by non-magnetic
V$^{5+}$O$_4$ tetrahedra shown in green.  The grey spheres
represent K$^+$ ions which separate the V$_3$O$_8$ 2d sheets.  (b)
Projection of the crystal structure perpendicular to the $c$-axis.
The intralayer coupling between S=1/2 V$^{4+}$ ions is shown by
blue arrows. (c) Projection of the crystal structure perpendicular
to the $a$-axis. The interlayer coupling, $J_c$ is shown by blue
arrows.} \label{structure}
\end{figure}

\begin{figure}
\centering
\includegraphics[width=\columnwidth,clip]{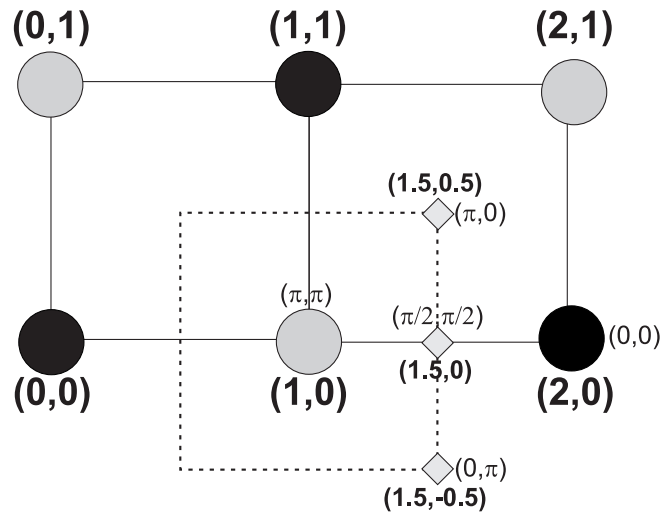}
\caption{Reciprocal space diagram for K$_2$V$_3$O$_8$.  Structural
Bragg reflections are indicated by black circles and magnetic
reflections by gray circles.  Several high symmetry zone boundary
points are indicated by gray diamonds.  All points are labelled by
the reciprocal space indices relevant to K$_2$V$_3$O$_8$ ($h$ and
$k$).  For reference, several points are additionally labelled
using the square lattice notation. As can be seen from this
notation, the square lattice is rotated by 45$^\circ$ from the
lattice of K$_2$V$_3$O$_8$.  The dashed lines show the
antiferromagnetic zone boundary around the (1,0) zone center.}
\label{zone}
\end{figure}

\section{K$_2$V$_3$O$_8$: General Properties}

K$_2$V$_3$O$_8$ crystallizes in the fresnoite structure with
tetragonal space group P4bm and lattice constants a=8.87 \AA~ and
c=5.2 \AA \cite{Galy}.  The room temperature structure, shown in
Fig. \ref{structure}, consists of vanadium oxide planes composed
of S=1/2, V$^{4+}$-O$_5$ square pyramids and nonmagnetic
V$^{5+}$-O$_4$ tetrahedra. The intralayer coupling constant, $J$,
is small due to the long V$^{4+}$-O-O-V$^{4+}$ superexchange path
as can be seen in Fig. \ref{structure}b which also shows the
square arrangement of the V$^{4+}$ ions. The 2d nature of the
magnetic system results from layers of K$^{+}$ ions which separate
the vanadium oxide layers. Powder magnetization measurements were
well described by the QSLHAF with a small near neighbor coupling
constant of 1.09 meV \cite{Liu}. This material is found to order
magnetically with T$_N$=4 K \cite{Lumsden} indicating a finite
interlayer coupling $J_c$ (Fig. \ref{structure}c).  The magnetic
structure in the absence of an applied magnetic field is a simple
two-sublattice antiferromagnetic arrangement with spins aligned
along the crystallographic $c$-axis with ferromagnetic stacking of
the 2d layers \cite{Lumsden}. A non-zero magnetic field results in
phase transitions when applied along both the $c$-axis and in the
tetragonal basal plane. The $c$-axis field induced phase
transition is a rather conventional spin-flop transition while the
presence of a basal plane applied field induces a peculiar spin
reorientation transition where the spins rotate from the easy
$c$-axis into the basal plane while remaining normal to the
applied field direction. These observations were described by the
following two-dimensional Hamiltonian which includes a
Dzyaloshinskii-Moriya (DM) interaction and an additional $c$-axis
anisotropy \cite{Lumsden},
\begin{eqnarray} \label{Hamiltonian}
  {\cal H} &=& J\sum_{\langle i,j \rangle}{\bf S}_i \cdot {\bf
S}_j+D_z\sum_{\langle i,j \rangle}\left({\bf S}_i\times {\bf S}_j\right)_z \nonumber \\
   &+& E_z\sum_{\langle i,j \rangle}S_{iz}S_{jz}+\sum_{i} {\bf H} \cdot {\bf S}_{i}.
\end{eqnarray}
where $\langle i,j\rangle$ represents near-neighbor pairs of
spins, and ${\bf S}_i$, ${\bf S}_j$ are the spin operators for
sites $i$,$j$. The applied field required to induce both the basal
plane spin reorientation and the $c$-axis spin flop transition
allowed for estimates of $E_z/J$ of 0.0012 and $D_z/J$ of 0.04.

Thermal conductivity measurements on K$_2$V$_3$O$_8$ indicate
dramatic enhancements in the low temperature heat transport in the
presence of an applied magnetic field \cite{Sales}.  These same
measurements indicate anomalies at 110 K \cite{Sales} suggesting a
phase transition at this temperature.  Optical measurements
clearly show anomalies in the local structure of the VO$_5$ square
pyramids on passing through 110 K \cite{Choi} indicating the
structural nature of this phase transition.

Recently, a more comprehensive theoretical treatment of the
magnetic properties of K$_2$V$_3$O$_8$ has been carried out
\cite{Chernyshev}. This work pointed out that weak ferromagnetism
induced by the DM interaction is ``hidden'' in K$_2$V$_3$O$_8$ in
the zero field limit by a dominant $c$-axis anisotropy. The
effects of the DM interaction are revealed in the presence of an
applied magnetic field and analytic expressions for the field
strength required to induce both the spin-flop and spin
reorientation transitions were provided.  The field dependence of
gap energies in the magnetic excitations were calculated and the
effects of quantum corrections were included. One interesting
theoretical prediction is a non-linear form for the gap energy as
a function of field applied in the basal plane.

The topic of this manuscript will be inelastic neutron scattering
measurements in the limit of zero applied magnetic field.  If we
set H=0 in Eq. \ref{Hamiltonian} and add a ferromagnetic coupling
between the 2d layers, as expected from the known zero-field
magnetic structure, linear spin-wave theory applied to the
resulting Hamiltonian yields two degenerate spin wave modes with
dispersion,
\begin{equation}\label{dispersion}
\omega_{\bf{Q}}=2\tilde{J}\sqrt{ \left(
1+\frac{E_z}{\tilde{J}}-\frac{J_c}{\tilde{J}}
\gamma_{\bf{Q_{\bot}}} \right )^2-\left[
1+\left(\frac{D_z}{\tilde{J}} \right)^2 \right ]
\gamma_{\bf{Q_{\|}}}^2}
\end{equation}
where
\begin{equation}
\gamma_{\bf{Q_{\|}}} = \cos(h\pi) \cos(k\pi);~
\gamma_{\bf{Q_{\bot}}}=\sin^2(l\pi).
\end{equation}
Here, we introduce an effective coupling constant,
$\tilde{J}=Z_{c}J$ where $Z_{c}$ represents the quantum
renormalization of the coupling constant with a best theoretical
estimate of $Z_{c}$=1.18 \cite{Igarashi1,Singh}. Note that as a
consistency check, if we set $J_c$=$E_z$=$D_z$=0, we recover the
result for the QSLHAF,
\begin{equation}
\omega_{\bf{Q}}=2\tilde{J}\sqrt{ 1-\cos^2(h\pi) \cos^2(k\pi)}.
\end{equation}
where the unit cell is rotated by 45$^\circ$ from the conventional
square lattice unit cell (see Fig. \ref{zone}).

Inelastic neutron scattering experiments measure a convolution of
the instrumental resolution function with the partial differential
cross-section,
\begin{equation}\label{crosssection}
\frac{d^2\sigma}{d \Omega d\omega} \sim \frac{k_i}{k_f}\mid F(Q)
\mid ^2 \left[n(\omega)+1\right] (1+\cos^2
\varphi)S(\bf{Q},\omega)
\end{equation}
where $k_i$ and $k_f$ are the magnitudes of the incident and final
neutron wavevectors, F(Q) is the magnetic form factor, in this
case for V$^{4+}$, $n(\omega)+1$ is the Bose occupation factor,
$(1+\cos^2 \varphi)$ is a polarization term reflecting the spin
direction ($\varphi$ is the angle between $\bf{Q}$ and the easy
$c$-axis) and $S(\bf{Q},\omega)$ is the dynamical structure
factor. Spin wave theory applied to the above Hamiltonian yields
\begin{equation}\label{sqw}
S( \mathbf{Q} , \omega) \sim \frac{\tilde{J}}{\hbar
\omega_{\bf{Q}}} \left( 1 + \frac{E_z}{\tilde{J}} -
\frac{J_c}{\tilde{J}} \gamma_{\bf{Q_{\bot}}} -
\gamma_{\bf{Q_{\|}}}\right) \delta(\omega-\omega_{\bf{Q}})
\end{equation}

\section{Experimental Details}

About 100 grams of V$_2$O$_5$ (99.995\%) and 76 grams of
K$_2$CO$_3$ (99.997\%) was loaded into a 250 ml Pt crucible and
slowly heated in air to 700$^{\circ}$C and held for 2 hours. The
powder was added in two stages because of substantial foaming. The
molten KVO$_3$ flux was then cooled to room temperature and the Pt
crucible and solidified KVO$_3$ loaded into a fused silica ampoule
that is necked down at the top. Using a long funnel, 9 grams of
VO$_2$ (99\%) was added to the Pt crucible and the entire ampoule
was evacuated and sealed. The sealed ampoule was loaded into a
furnace, heated to 850 $^{\circ}$C for 6 hours, cooled to 700 °C
over 1h and then cooled to about 400 $^{\circ}$C at 1
$^{\circ}$C/h, followed by furnace cooling to room temperature.
The KVO$_3$ flux was then removed with a combination of warm water
and ultrasonic vibration. The resulting K$_2$V$_3$O$_8$ crystals
are black rectangular plates with typical dimensions of 5 x 5 x 1
mm$^3$.

\begin{figure}
\centering
\includegraphics[angle=0,width=\columnwidth,clip]{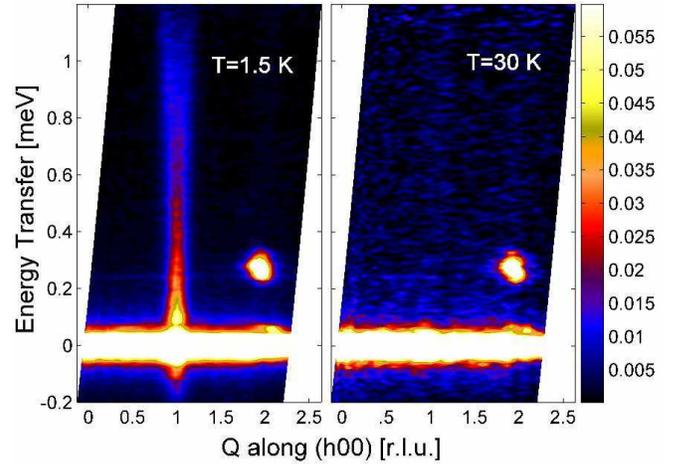}
\caption{(Color online) Low energy magnetic excitation spectrum of
K$_2$V$_3$O$_8$ at temperatures of 1.5K and 30K.  The emergence of
the excitations on cooling below the peak in the susceptibility
clearly shows the magnetic nature of the excitations.  The
temperature independent feature at H=1.8 and E=0.25 meV is
spurious and is related to the presence of the nearby (2,0,1)
structural Bragg reflection. The color bar indicates intensity of
scattered neutrons in arbitrary units.}
\label{iris_low_energy_contour}
\end{figure}

\begin{figure}
\centering
\includegraphics[angle=0,width=\columnwidth,clip]{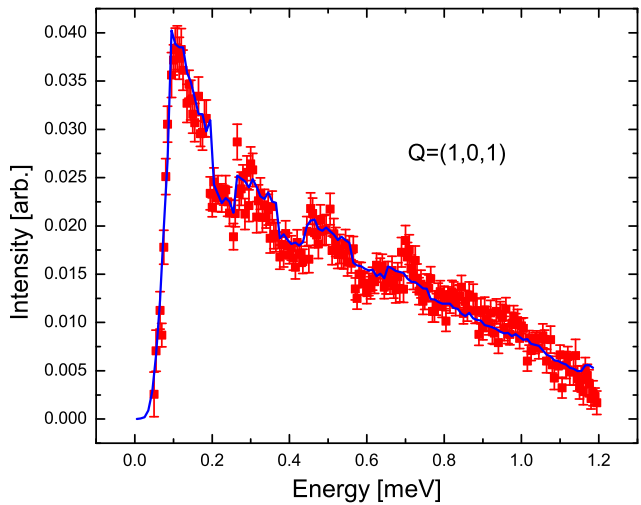}
\caption{(Color online) Cut along the energy axis for H ranging
from 0.9 to 1.1 showing the clear presence of a small energy gap
of about 70 $\mu$eV near the (1 0 1) antiferromagnetic zone
center.  The solid line represents the cut through simulated data
resulting in extremely good quantitative agreement between the
data and the model.  The strange lineshape is a result of the
small number of detectors which cut through the dispersion at
discrete positions.} \label{edependent_cut}
\end{figure}

Inelastic neutron scattering experiments on single crystal samples
were performed using the backscattering spectrometer, IRIS at the
ISIS neutron scattering facility, the RITA II triple-axis
spectrometer at the Paul-Scherrer Institut and the SPINS
triple-axis spectrometer at the NIST Center for Neutron Research.
The IRIS and RITA II experiments were performed on an array of
$\sim$ 40 single crystals visually co-aligned with a total mosaic
spread of about 4$^\circ$.  Improvements in crystal size allowed
the SPINS experiment to be performed on an array of 5 single
crystals which were much more accurately co-aligned with a final
mosaic spread of about 0.5$^\circ$.  For all experiments, the
crystals were mounted on an aluminum sample mount which was
connected to the cold finger of a $^4$He cryostat with a minimum
temperature of about 1.5 K.  The IRIS experiments used PG (002) to
select a fixed E$_f$ of 1.845 meV yielding a resolution of
$\sim$18 $\mu$eV at the elastic position. Measurements were
performed using two sets of chopper settings allowing for studies
with energy transfers covering a range from -0.2 to 1.2 meV and
0.5 to 2.8 meV respectively. Both the RITA II and SPINS
triple-axis spectrometers used PG (002) as the monochromator with
pre-monochromator collimation determined by the presence of a
neutron guide and an 80' post-monochromator collimator.  The
majority of measurements were performed with a fixed final energy
of 5 meV and a cooled Be filter before the analyzer resulting in
an energy resolution of $\sim$0.25 meV at the elastic position.
Both instruments employ a multiblade PG (002) analyzer
configuration and measurements were made with this analyzer array
either in a flat or focussing configuration.

For means of reference, we have included a map of reciprocal space
in Figure \ref{zone} with nuclear zone centers indicated by black
circles, antiferromagnetic zone centers by gray circles, and
several high symmetry antiferromagnetic zone boundary points by
gray diamonds. To allow for easier comparison with previous
measurements, the relevant region of reciprocal space is labelled
by both $h$ and $k$ for K$_2$V$_3$O$_8$ together with the labels
for the conventional square lattice.  This labelling clearly shows
that the unit cell of K$_2$V$_3$O$_8$ is rotated by 45$^\circ$
from the square lattice unit cell.

\section{Results: Long Wavelength Excitations}

One clear prediction of the dispersion shown in Eq.
\ref{dispersion} is the presence of an energy gap at the
antiferromagnetic zone center of,
\begin{equation}\label{gap}
\Delta=2\tilde{J}\sqrt{\left( 1+ \frac{E_z}{\tilde{J}}\right)^2 -
\left[ 1+\left( \frac{D_z}{\tilde{J}} \right)^2\right]}
\end{equation}
Using the previously determined experimental values for $J$,
$E_z/J$, and $D_z/J$ \cite{Lumsden}, together with the expected
quantum correction $\tilde{J}$=1.18$J$ \cite{Igarashi1,Singh} we
expect this energy gap to be about 73 $\mu$eV. In order to explore
the low energy region of the dispersion, we performed initial
measurements on the IRIS backscattering spectrometer.  For these
measurements, the sample was mounted in the ($h$ 0 $l$) scattering
plane. Measurements of the low energy portion of the magnetic
excitation spectrum at temperatures of 1.5 K and 30 K are shown in
Figure \ref{iris_low_energy_contour}. This plot shows the
inelastic intensity as a color map as a function of component of
momentum transfer along the ($h$ 0 0) direction and energy
transfer. The excitations, clearly seen in the T=1.5 K data, are
found to vanish at higher temperatures providing clear evidence of
their magnetic nature. The temperature independent bright
intensity spot observed near $h$=1.8 and E=0.25 meV is a spurious
feature, not uncommon in an inverse geometry, time-of-flight
spectrometer where a white beam is incident on the sample, related
to a nearby (2,0,1) structural Bragg reflection.

%\begin{figure}
%\centering
%\includegraphics[angle=-90,width=\columnwidth,clip]{Figures/simulated_scattering}
%\caption{Simulation of the expected scattering obtained by
%convolving $S(\textbf{Q},\omega)$ as defined in Eq. \ref{sqw} with
%the instrumental resolution function.  Qualitative comparison with
%the measured excitation spectrum shown in Fig.
%\ref{iris_low_energy_contour} suggests that the model accurately
%describes the experimental observations.}
%\end{figure}

\begin{figure}
\centering
\includegraphics[angle=0,width=\columnwidth,clip]{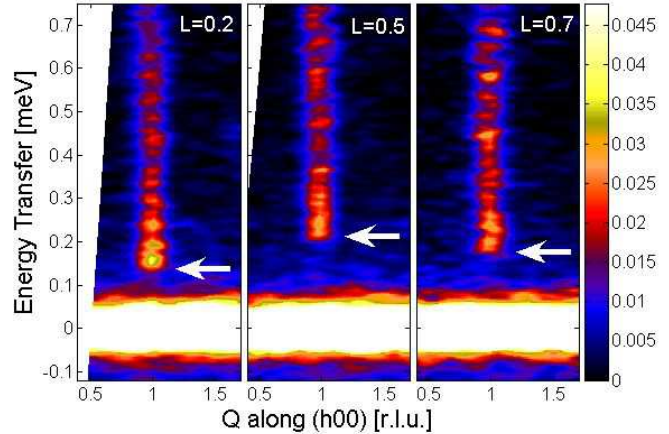}
\caption{(Color online) Measured excitation spectrum at several
different sample rotations showing the dispersion along the
crystallographic c-axis between the planes.  The color bar
represents the intensity of scattered neutrons in arbitrary
units.} \label{caxis_contour}
\end{figure}

\begin{figure}
\centering
\includegraphics[angle=0,width=\columnwidth,clip]{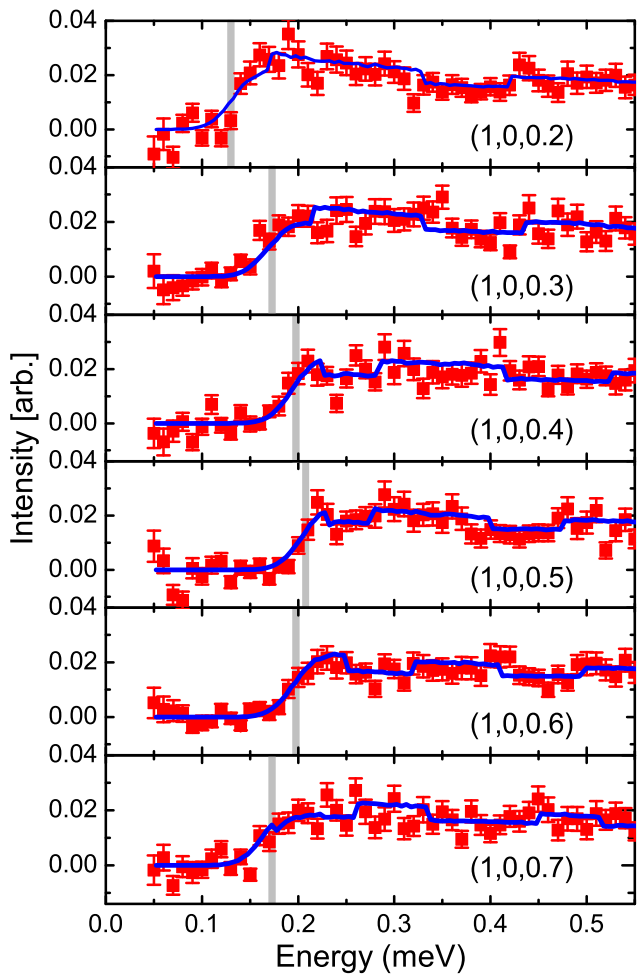}
\caption{(Color online) Comparison of the measured data at several
different values of L with the simulated scattering.  The gray
bars represent the dispersion at the respective L values. This
results in a ferromagnetic interplanar coupling of
$J_c/\tilde{J}$=-0.0028$\pm$0.0005 meV showing the nearly 2d
nature of the magnetic system.} \label{caxiscut}
\end{figure}

Careful examination of Figure \ref{iris_low_energy_contour}
suggests a slight reduction in inelastic intensity as the energy
approaches the elastic position which may be an indication of a
gap in the excitation spectrum.  To see this more clearly, an
energy dependent background was subtracted from the data shown in
Figure \ref{iris_low_energy_contour} and an energy dependent cut
through this background subtracted data for \emph{h} ranging from
0.9 to 1.1 is shown in Figure \ref{edependent_cut}.  This plot
clearly shows a gap in the excitation spectrum which appears to
about 100 $\mu$eV.  The solid line shown in Figure
\ref{edependent_cut} represents the cross-section of Eq.
\ref{crosssection} convoluted with the instrumental resolution
function of the IRIS spectrometer.  The line is in excellent
agreement with the measured data and allows extraction of a zone
center energy gap of 72$\pm$9 $\mu$eV. As can be clearly seen from
Eq. \ref{gap}, the value of the gap cannot uniquely define both
$E_z/\tilde{J}$ and $D_z/\tilde{J}$. If we arbitrarily fix
$D_z/\tilde{J}$ to value of 0.04 \cite{Lumsden} and use a value of
$\tilde{J}$ of 1.275 (as will be shown later), we extract a value
of the $c$-axis anisotropy of $E_z/\tilde{J}$=0.0012$\pm$0.0001
consistent with the value determined previously from the locations
of the field-dependent phase transitions \cite{Lumsden}. On a
technical note, the apparent jumps in the data shown in Figure
\ref{edependent_cut} result from the small number of detectors
present on IRIS and these jumps are quite accurately reproduced in
the simulated scattering, shown by the solid line in Figure
\ref{edependent_cut}, when the instrumental resolution and
detector configuration are properly taken into account.

The same experimental configuration on IRIS can also be used to
measure the dispersion perpendicular to the 2d planes.  This is
accomplished by measuring the same spectrum at several different
sample rotations and the results are shown for $l\approx$0.2,0.5
and 0.7 in Figure \ref{caxis_contour}.  Note that these
designations for $l$ are only valid in the elastic position and
$l$ actually varies across the contour plot.  Nonetheless, the
dispersion along the $c$-axis can clearly be seen in this plot as
a variation in the gap represented by the arrows in Figure
\ref{caxis_contour}.  To quantitatively extract the dispersion
along the $c$-axis, the same cut along energy (for $h$ ranging
from 0.9 to 1.1) at several different sample angles is shown in
Figure \ref{caxiscut}.  The solid lines shown in the figure
represent a numerical convolution of the cross-section (Eq.
\ref{crosssection}) with the instrumental resolution.  Once again,
we see good agreement between the measured data and the simulation
for all sample rotation angles measured.  The grey vertical bars
shown in each panel represent the location of the energy gap for
the appropriate $l$-value clearly showing the $c$-axis dispersion.
The fits allow us to extract a value of the interplanar coupling
of $J_c/\tilde{J}$=-0.0028$\pm$0.0005 reflecting the two
dimensionality of the material.  Note that the negative sign
indicates the ferromagnetic nature of the interaction consistent
with the observed magnetic structure \cite{Lumsden}.

\begin{figure}
\centering
\includegraphics[angle=0,width=\columnwidth,clip]{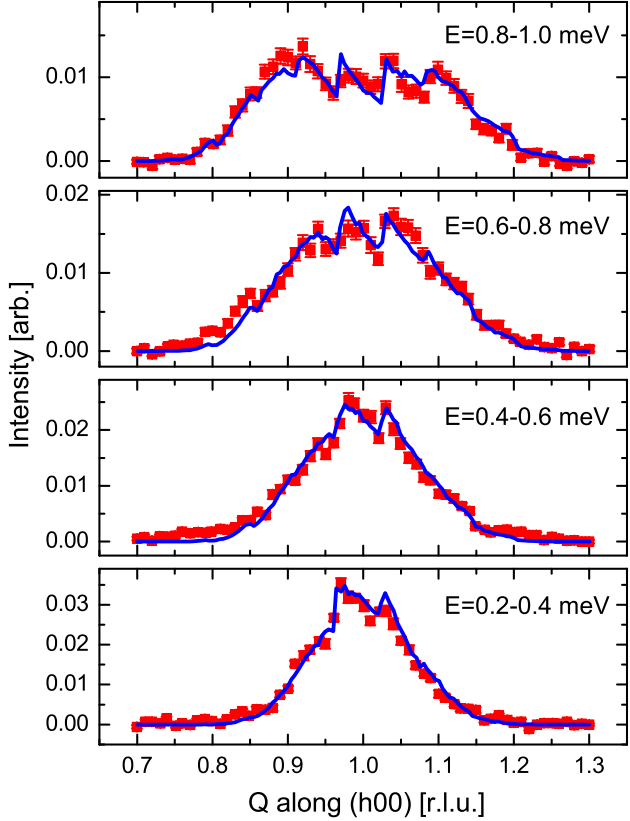}
\caption{(Color online) Comparison of the data and simulated
scattering at several ranges of energies.  As before, the model
can be seen to agree with the data up to 1 meV.  The limited Q
resolution of IRIS makes it difficult to resolve the dispersing
modes.} \label{highercuts}
\end{figure}

While IRIS is an ideal instrument for measuring features
associated with the energy gap, due to extremely good energy
resolution, it has Q-resolution limitations resulting from the
small number of detectors on the instrument and the large angle
accepted by each analyzer crystal each of which scatters into a
single detector. To see this, we plot cuts along the (\emph{h}00)
direction for several ranges of energy transfer in Figure
\ref{highercuts} together with the corresponding numerical
convolution of the expected cross-section (Eq. \ref{crosssection})
with the instrumental resolution function. While the simulation is
seen to agree well with the measured data, it is extremely
difficult to resolve the excitations and, hence, to extract
physical parameters from these fits due to limited Q resolution.

\begin{figure}
\centering
\includegraphics[angle=0,width=\columnwidth,clip]{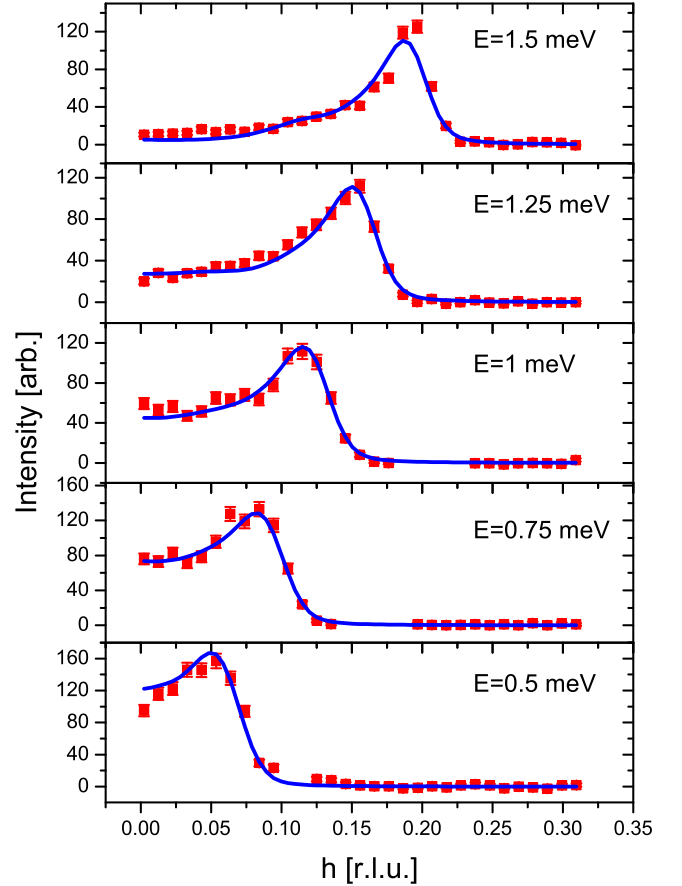}
\caption{(Color online) Constant-E scans measured using the SPINS
triple-axis spectrometer for energies ranging from 0.5 meV to 1.5
meV together with fits to S(\textbf{Q},$\omega$) convolved with
the instrumental resolution function.} \label{spins_constant_E}
\end{figure}

To extract information about the coupling constant, measurements
were performed on the SPINS and RITA II triple-axis spectrometers.
The multiblade analyzer system was configured in a horizontally
focussing  configuration for these measurements which increases
intensity while maintaining good energy resolution at the
sacrifice of Q resolution in the direction perpendicular to $k_f$.
The results shown in Figure \ref{spins_constant_E} represent
constant-E scans performed with the sample mounted in the
($h$,$k$,0) plane using the SPINS triple-axis spectrometer.  These
measurements were performed along the ($h$,1,0) direction which
corresponds to the direction from ($\pi$,$\pi$) (for \emph{h}=0)
toward the antiferromagnetic zone boundary, ($\pi$/2,$\pi$/2) (for
\emph{h}=0.5). The much better Q resolution of the triple-axis
spectrometer can be seen from the clearly dispersive mode in the
data shown in Fig. \ref{spins_constant_E}. The use of the
horizontally focussing configuration made analysis of the data
complicated in this case. Typically, this configuration is used
with an irrelevant direction in the scattering plane which can be
oriented along the direction perpendicular to $k_f$. However, in
these measurements, both directions within the ($h$,$k$,0) plane
are strongly dispersive. This complicated the analysis as typical
Gaussian approximations to the resolution function were not able
to reproduce the sharp features observed in the data.
Consequently, the cross-section appropriate to K$_2$V$_3$O$_8$
(Eq. \ref{crosssection}), was convolved with the resolution
function for each of the 7 individual analyzer blades and the
contribution from each blade was added with the correct phase to
yield the solid lines shown in Figure \ref{spins_constant_E}. To
correctly duplicate the effective collimation introduced by the
small width of the individual analyzer blades, the more rigorous
Popovici approximation \cite{Popovici} to the triple-axis
resolution function was required. Several scans ranging from 0.5
meV to 1.5 meV were simultaneously fit with a single overall
amplitude clearly showing that intensity modulation represented by
$S(\bf{Q}, \omega)$ (Eq. \ref{sqw}) provides an excellent
description of the data.
%These fits resulted
%n an overall reduced $\chi^2$ of 3.26 over 225 individual data
%points taken from 11 individual scans
This analysis allowed extraction of the 2d coupling constant
$\tilde{J}$ of 1.275$\pm$0.03 meV.

\begin{figure}
\centering
\includegraphics[angle=0,width=\columnwidth,clip]{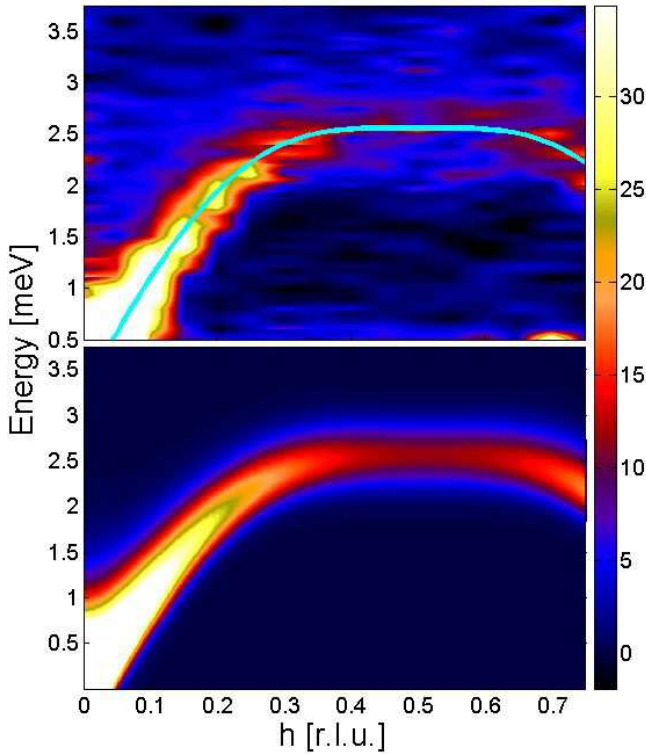}
\caption{(Color online) Summary of measurements performed along
the (h,1+h,0) direction using the RITA II triple-axis spectrometer
is shown in the upper panel.  The lower panel shows the result of
the best fit simulation to S(\textbf{Q},$\omega$) convolved with
the instrumental resolution function.  The color bar represents
the intensity of scattered neutrons in counts per $\sim$2
minutes.} \label{psi_h1h0}
\end{figure}

%\begin{figure}
%\centering
%\includegraphics[angle=0,width=\columnwidth,clip]{Figures/ZBapproach}
%\caption{Add in some text here.} \label{zbcontour}
%\end{figure}
%\begin{figure}
%\centering
%\includegraphics[angle=0,width=\columnwidth,clip]{Figures/zbplots_psi}
%\caption{Add in some text here.} \label{zbdisppsi}
%\end{figure}

As mentioned previously, the coupling constant, $\tilde{J}$, is
written as an effective coupling and theoretical studies indicate
that quantum corrections to spin-wave theory result in an overall
renormalization of the dispersion by a scaling factor $Z_c$ such
that $\tilde{J}$=$Z_c J$ \cite{Igarashi1,Singh}.  As a consistency
check, we can calculate the value of $Z_c$ by comparing the
coupling constant above to the value of $J$=1.09 meV obtained from
magnetization measurements \cite{Liu} where quantum effects were
taken into account.  This results in $Z_c$=1.174$\pm$0.03, in
excellent agreement with the best theoretical estimate of $Z_c$ of
1.18 \cite{Igarashi1,Singh} justifying the designation of
K$_2$V$_3$O$_8$ as a good example of a QSLHAF. Using the
theoretical value for $Z_c$, we can extract the true coupling
constant, $J$, for K$_2$V$_3$O$_8$ of 1.08$\pm$0.03 meV.

\begin{figure}
\centering
\includegraphics[angle=0,width=\columnwidth,clip]{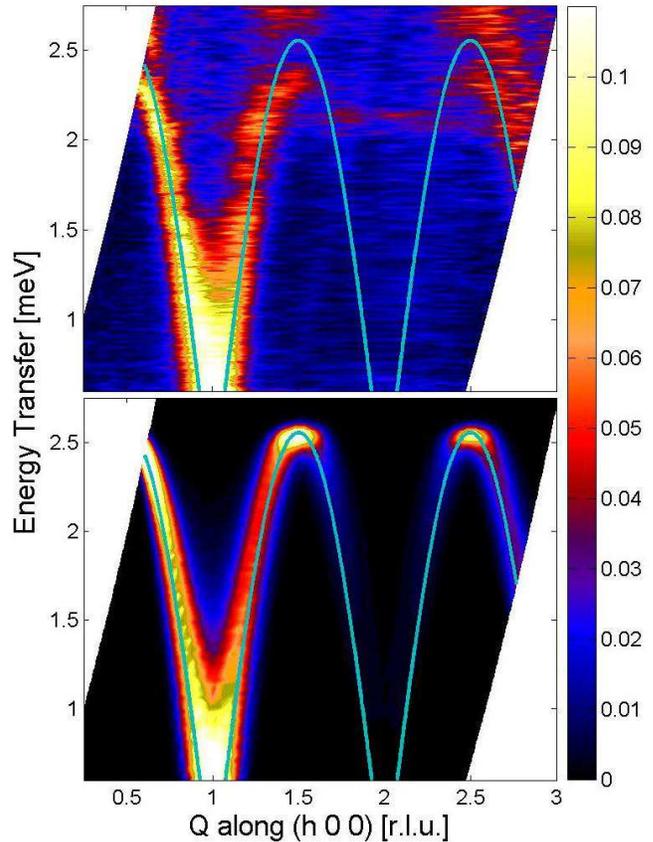}
\caption{(Color online) The measured dispersion from energy
transfers ranging from 0.5 to 2.8 meV using the IRIS
backscattering spectrometer is shown in the upper panel.  The
lower panel represents the results of a convolution of the
instrumental resolution function with the expected cross-section.
In both panels, the identical classical spin-wave dispersion
dispersion, corresponding to a $\tilde{J}$=1.275 meV is
superposed.  The color bar represents the intensity of scattered
neutrons in arbitrary units.} \label{iris_high_energy}
\end{figure}

Measurements of the dispersion were additionally performed along
the square lattice direction ($\pi$,$\pi$) to ($\pi$,0)
corresponding to the ($h$,1+$h$,0) direction in K$_2$V$_3$O$_8$.
Measurements along this direction were performed using the RITA II
triple-axis spectrometer with the sample mounted in the
($h$,$k$,0) scattering plane and the results are shown in the
upper panel of Figure \ref{psi_h1h0}.  The solid line in the upper
panel represents the classical spin wave dispersion with
$\tilde{J}$=1.275 meV which clearly describes the data well.  The
lower panel represents the best fit to a convolution of Eq.
\ref{crosssection} with the instrumental resolution function with
the dispersion represented by the solid line in the upper panel.
The fits agree very well with the data across the entire zone
except in the immediate vicinity of the zone boundary where the
peak intensity appears weaker than the calculated peak value.  The
behavior near the zone boundary will be discussed in more detail
in the next section.

\section{Results: Short Wavelength Behavior}

As mentioned in the introduction, the dynamical structure factor
of the QSLHAF only deviates from predictions of spin-wave theory
near the antiferromagnetic zone boundary. Consequently, to explore
the behavior in the vicinity of the zone boundary more carefully,
measurements on IRIS were extended with the choppers configured to
select a range of energy transfers from 0.5 to 2.8 meV and the
resulting data are shown in the upper panel of Figure
\ref{iris_high_energy}.  As can be clearly seen from the data, we
observe well defined excitations up to the zone boundary. However,
as can be seen from the solid line in the upper panel, the
dispersion curve which describes the data well near the zone
center seems to disagree with the data near the zone boundary.  In
the lower panel of Figure \ref{iris_high_energy}, we show the
result of a convolution of the instrumental resolution with the
expected cross-section together with the dispersion curve used to
generate the simulation.  In comparing this to the measured data,
it is also clear that once the amplitude is set to provide good
agreement with the data near the zone center, the calculated
intensity near the zone boundary exceeds that of the measurement.
However, closer examination of the measured intensity indicates
the presence of two modes with intensity in the second mode only
appearing near the zone boundary.

\begin{figure}
\centering
\includegraphics[angle=0,width=\columnwidth,clip]{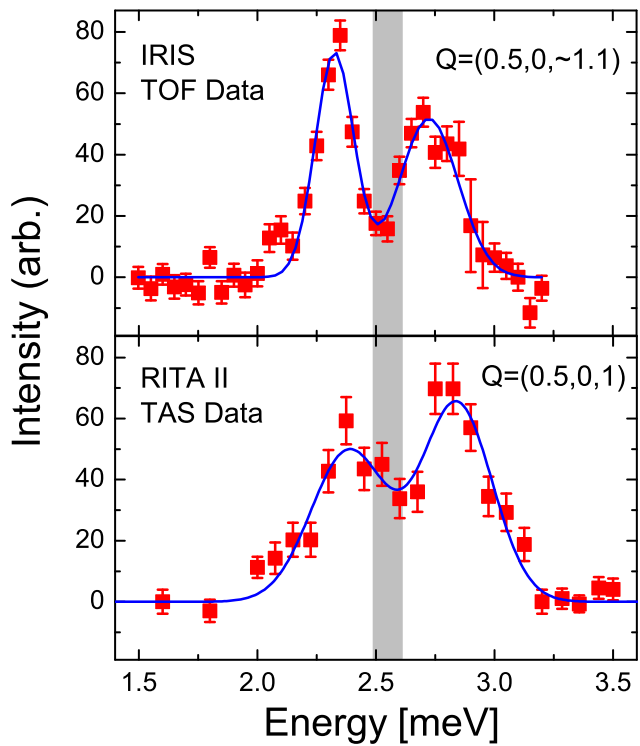}
\caption{(Color online) Cut through the IRIS data for a range of
$h$ from 0.4 to 0.6 as a function of energy transfer with a
nominal \textbf{Q} of (0.5,0,1.1) is shown in the upper panel. The
lower panel shows the corresponding constant-Q scan through
(0.5,0,1) using the RITA II triple-axis.  The solid gray bar shows
the range of zone boundary energies resulting from measurements
closer to the zone center and the predictions of linear spin-wave
theory.} \label{comparezb}
\end{figure}

Instrumentally, the IRIS spectrometer exhibits various spurious
scattering features around 3 meV energy transfer.  The presence of
these features led to scepticism as to the validity of the
observed two modes near the zone boundary.  To confirm this
observation, measurements were performed in the identical
($h$,0,$l$) scattering plane using the RITA II triple-axis
spectrometer.  The instrument was configured with E$_f$=4.6 meV
and the resulting constant-Q scan at \textbf{Q}=(0.5,0,1) is shown
in the lower panel of Figure \ref{comparezb}.  In the upper panel,
we show a cut through the IRIS data from $h$=0.4 to $h$=0.6 with
the sample slightly rotated from the measurement shown in Figure
\ref{iris_high_energy} so as to emphasize the data near $h$=0.5.
Both the time-of-flight and triple-axis data clearly show the
presence of two peaks near the zone boundary.  For reference, the
gray box represents the predicted zone boundary energy
corresponding to $\tilde{J}$=1.275 meV with the width of the box
corresponding to the uncertainty of $\pm$0.03 meV. Clearly, in
both measurements, the calculated value of the zone boundary
energy falls between the two measured peaks. Note that the exact
energies of the two modes differ slightly between the two
instruments, a discrepancy which can be accounted for by
differences in energy calibration.

\begin{figure}
\centering
\includegraphics[angle=0,width=\columnwidth,clip]{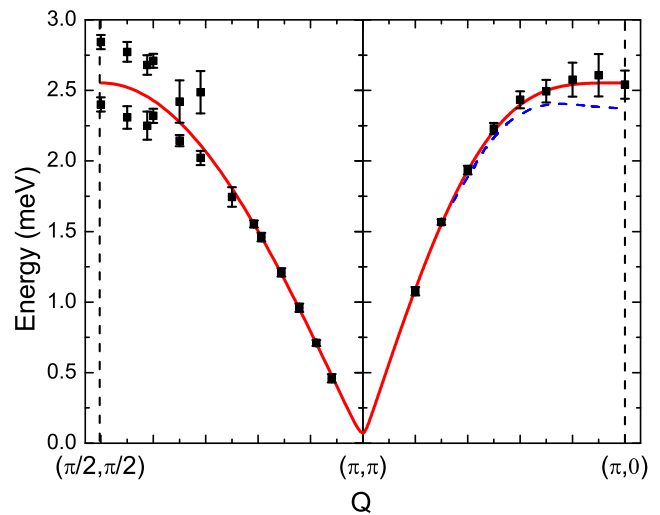}
\caption{(Color online) Summary of the full triple-axis measured
dispersion. The measurements are the result of experiments
performed at both RITA II and SPINS for experiments in both the
($h$,$k$,0) and ($h$,0,$l$) scattering planes.  The solid line
corresponds to a fit to linear spin-wave theory for data near the
zone center.  The dashed line represents the quantum corrections
to the dispersion \cite{Ronnow}.} \label{finaldisp}
\end{figure}

The full dispersion, measured using triple-axis spectrometers
along high symmetry directions in the 2d plane, is summarized in
Figure \ref{finaldisp} with \textbf{Q} plotted in square lattice
notation. The data shown in this figure are the result of
experiments performed using both the RITA II and SPINS triple-axis
spectrometers with the sample oriented in both the ($h$,$k$,0) and
($h$,0,$l$) scattering planes.  The measured data along the
($\pi$,$\pi$) - ($\pi$,0) direction and near the zone center along
the ($\pi$,$\pi$) - ($\pi$/2,$\pi$/2) were fit to the expected
S(\textbf{Q},$\omega$) convolved with the instrumental resolution
function. The amplitude, coupling constant, and a Lorentzian
broadening term were allowed to vary for each of these data
points.  As this model no longer describes the data well near the
($\pi$/2,$\pi$/2) zone boundary point, the data here was fit to
Gaussian lineshapes with widths constrained to be broader than
that expected from resolution.  The solid line is the expected
classical dispersion (Eq. \ref{dispersion}) with
$\widetilde{J}$=1.275 meV, D$_z$/$\widetilde{J}$=0.04,
E/$\widetilde{J}$=0.0012, and J$_c$/$\widetilde{J}$=-0.0028. This
solid line is seen to agree well with the data near the zone
center but deviates on approaching the ($\pi$/2,$\pi$/2) zone
boundary point. As the data begins to deviate from the solid line,
an additional mode appears. As reflected by the size of the error
bars, this mode is rather weak when it first appears and gains
intensity on approaching the zone boundary. As shown previously in
Figure \ref{comparezb}, the calculated dispersion falls between
the two modes on approaching ($\pi$/2,$\pi$/2). There is no clear
evidence for two peaks near the ($\pi$,0) zone boundary point.
However, as indicated by the large error bars near this point, the
measured spectrum is broader than would be expected from
resolution.  The dashed line represents the expected quantum
dispersion \cite{Ronnow}.  While it does appear that the data
agrees better with the classical dispersion, there very well may
be two modes near the ($\pi$,0) zone boundary point as well.  As
such, the points shown on the dispersion may represent the mid
point of the two modes the lower of which would be quite close to
the quantum dispersion.

\begin{figure}
\centering
\includegraphics[angle=0,width=\columnwidth,clip]{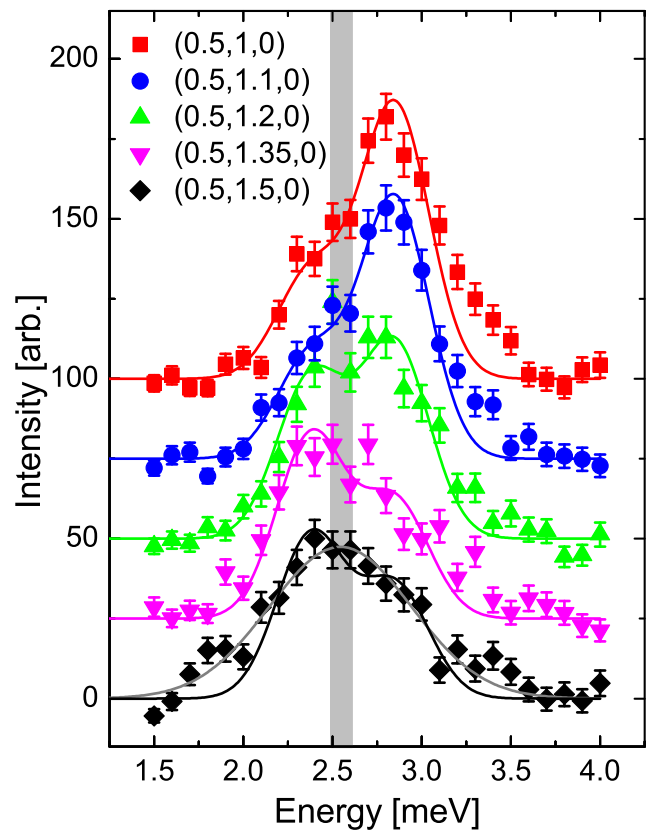}
\caption{(Color online) Constant-Q scans along the zone boundary
for Q values ranging from (0.5,1,0) to (0.5,1.5,0).  The solid
lines present fits to two Gaussians with positions fixed at values
measured in higher resolution conditions with fixed widths and
only the amplitudes varying at different Q positions.  The gray
solid line through the black data shows a fit to a single,
broadened Gaussian.  The solid gray bar shows the range of values
of zone boundary energies predicted using spin-wave theory and
consistent with measurements near the zone center.}
\label{zbdispspins}
\end{figure}

As mentioned in the introduction, the one region where
measurements and quantum calculations on the QSLHAF deviate from
the predictions of spin-wave theory is in measuring the zone
boundary energy along the direction ($\pi$/2,$\pi$/2) to
($\pi$,0). Consequently, we have measured the excitations along
this direction using the SPINS triple-axis spectrometer to study
the evolution of the observed two modes and the result is shown in
Figure \ref{zbdispspins}. For reference, as shown in Figure
\ref{zone}, \textbf{Q}=(0.5,1,0) corresponds to the square lattice
point ($\pi$/2,$\pi$/2) while \textbf{Q}=(0.5,1.5,0) corresponds
to ($\pi$,0).  The presence of two modes can also be seen in this
scattering plane.  To further emphasize this, the solid lines in
the figure are the results of two Gaussians centered at the
positions measured under higher resolution conditions, shown in
Fig. \ref{comparezb}, with fixed widths and only the amplitudes of
the Gaussians varying. The results clearly show an intensity
distribution which varies across the zone boundary with the higher
energy mode having its greatest intensity near (0.5,1,0) and
continuously decreasing on approaching (0.5,1.5,0). In fact,
although this two mode model still describes the data fairly well
near (0.5,1.5,0), a better description, as shown by the solid gray
line is that of a broadened Gaussian. This is consistent with the
previously presented data showing the dispersion along the
($h$,1+$h$,0) direction (Fig. \ref{psi_h1h0} and \ref{finaldisp})
which suggests only a single mode along this direction albeit with
increased width near the zone boundary. As mentioned previously,
there may be two modes in this direction as well and the weaker
intensity of the upper mode may act to increase the difficultly in
resolving the individual modes. For reference, the solid gray bar
in Figure \ref{zbdispspins} shows the range of values for the zone
boundary energy given by the dispersion of Eq. \ref{dispersion}
with $\widetilde{J}$=1.275 meV. The width of this bar is
reflective of the uncertainty in the predicted classical zone
boundary energy and is approximately 5\%. Theoretically,
dispersion between ($\pi$/2,$\pi$/2) and ($\pi$,0) is predicted to
be 7-9\%, less than twice the width of the bar. Clearly, the
measured data near the zone boundary is far too broad to allow
observation of such a small zone boundary dispersion and the
presence of the unexpected two modes makes such an investigation
meaningless.

\section{Origin of Splitting}

\subsection{Magnon-Phonon Interaction}

The observation of two modes near the zone boundary is clearly not
expected from the Hamiltonian postulated for K$_2$V$_3$O$_8$ (Eq.
\ref{Hamiltonian}).  We have experimentally characterized the zone
boundary properties quite extensively.  The higher energy mode
only seems to have significant spectral weight in the immediate
vicinity of the zone boundary.  We have tried fitting the data
closer to the zone center using a model incorporating two
dispersive modes and no significant improvement in fit quality was
observed supporting the notion that the high energy mode is only
significant near the zone boundary.  In addition, the presence of
two modes is much clearer near \textbf{Q}=(0.5,1,0) (or
($\pi$/2,$\pi$/2) in square lattice notation) and the intensity of
the upper mode falls off continuously in moving away from this
position along the zone boundary.  Eventually, this evolves into
what is best described as a broadened single peak at
\textbf{Q}=(0.5,1.5,0) (($\pi$,0) in square lattice notation). In
addition, both modes vanish upon increasing the temperature to
$\sim$30 K indicating that both are of magnetic origin.  This
rules out the possibility of a low-lying phonon mode interacting
with the spin waves.  Any proposed model to describe the observed
scattering must satisfy these experimental observations.

\subsection{Additional Magnetic Interactions}

Although we haven't determined a single, consistent explanation
for the peculiar zone boundary properties of K$_2$V$_3$O$_8$,
there are a number of possible explanations which we can discuss.
Of course, the first possibility is that of additional couplings
in the Hamiltonian. Longer range interaction do, in fact, tend to
manifest themselves near the zone boundary \cite{Coldea}. Although
we can't rule out such a possibility, it is difficult to conceive
of a coupling term which will only produce intensity near the zone
boundary.
%The majority of
%additional coupling terms have a stronger effect near the zone
%center (for instance, a split energy gap) and, typically, changes
%near the zone boundary are much weaker as the energy there is
%considerably larger.
The particular case of next-near-neighbor coupling has been
studied quite extensively using both linear spin-wave theory
\cite{Merino} and series expansion techniques \cite{Zheng2}.
Although the inclusion of such a coupling term does modify the
dispersion, with the largest effects observed near the zone
boundary, no evidence for a split excitation spectrum is predicted
\cite{Merino,Zheng2}. The symmetry arguments presented by Bogdanov
\emph{et~al.} \cite{Bogdanov} suggest that K$_2$V$_3$O$_8$
together with Ba$_2$CuGe$_2$O$_7$ \cite{Zheludev1,Zheludev2} are
members of an interesting class of compounds where chiral
inhomogeneous magnetic structures can coexist with weak
ferromagnetism.  The net result of this symmetry analysis is the
existence of an additional, symmetry allowed term in the
Hamiltonian describing K$_2$V$_3$O$_8$. The form of this chiral
coupling term is rather complicated and its effects on the
excitation spectrum are unclear at this point indicating the need
for further theoretical investigations.

\subsection{Multimagnon Scattering}

Another possible explanation for the additional scattering
observed at the zone boundary is multi-magnon scattering.  A
recent theoretical investigation \cite{Syljuasen} based on the
assumption that high-energy fermions exist in the QSLHAF suggest a
very structured multi-magnon continuum resulting in a rather
sharply peaked extra mode around the ($\pi$/2,$\pi$/2) point and
much broader scattering around ($\pi$,0) in qualitative agreement
with our experimental observations.  However, the strength of the
scattering in the additional mode makes such an interpretation
very unlikely.  Theoretical studies using both series expansion
\cite{Zheng} and quantum Monte Carlo \cite{Sandvik} indicate a
significant amount of spectral weight in the multi-magnon
spectrum.  The calculated ratio of the longitudinal to transverse
structure factors for unpolarized neutrons is 29\% (34\%) at
($\pi$,0) while at ($\pi$/2,$\pi$/2) the ratio is 31\% (35\%)
where the values quoted are the results of series expansion
\cite{Zheng} (quantum Monte Carlo \cite{Sandvik}) studies.
Although these values are rather large reflecting the strength of
the multi-magnon scattering expected for the QSLHAF, they are
somewhat deceiving in that they assume that neutron scattering
will fully measure two transverse modes and one longitudinal mode.
In reality, the neutron only sees components of magnetic moment
normal to the wavevector transfer resulting in a correction term
which has the form $1+\cos^2(\varphi)$ for the two transverse
modes and $\sin^2(\varphi)$ for the longitudinal mode where
$\varphi$ is the angle between \textbf{Q} and the spin direction,
in this case the $c$-axis.  For the results presented in Figure
\ref{comparezb} the angle between the (0.5,0,1) direction and the
$c$-axis is about 30 degrees resulting in a correction term of
1.75 for the two transverse modes and 0.25 for the longitudinal
mode.  This implies, at this wavevector, that instead of seeing
about 30\% of the intensity, as predicted for the calculation
under the assumption of a 1:2 ratio of longitudinal to transverse
we would actually see 30\%*2*0.25/1.75 $\approx$ 8.5\%. Clearly,
the measurement at (0.5,0,1) shows intensities for the two modes
which are very similar in magnitude and is not consistent with the
upper mode being 8.5\% of the lower mode. This essentially
eliminates multi-magnon scattering as a source for the additional
mode.

\subsection{Orbital Effects}

Another possible source for the observed mode splitting is the
influence of orbital degrees of freedom.  Examination of the
dispersion curve (Fig. \ref{finaldisp}) is qualitatively
consistent with the behavior expected if a flat mode were
interacting with the spin wave dispersion resulting in repulsion
of the modes.  In addition, a flat excitation with a weak neutron
structure factor which was enhanced through mixing with the
spin-wave mode could explain the observed intensity modulation.
One potential source of such a flat mode would be a low-lying
crystal field excitation. A related effect was predicted
theoretically using a generalized Holstein Primakoff
transformation of the magnetic excitations in a system with spin
and orbital degrees of freedom in the presence of orbital
degeneracy \cite{Joshi}.  The net result of inclusion of orbital
degrees of freedom is an excitation spectrum with modes of pure
spin, pure orbital, and mixed spin-orbital character \cite{Joshi}.
Mixing is predicted between the spin and spin-orbital modes the
strength of which is greatest near the zone boundary and weaker
near the zone center. The scenarios postulated above require an
understanding of the crystal-field ground state together with the
approximate energies for the lowest excitations from this ground
state.

Symmetry analysis of the local square-pyramidal environment of the
V$^{4+}$ ion yields a splitting of the octahedral $t_{2g}$ levels
into a singlet ($d_{xy}$) and a doublet ($d_{xz}$, $d_{yz}$) and
the octahedral $e_g$ states into two singlets ($d_{x^2-y^2}$ and
$d_{z^2}$).  Consequently, the only possibility of an orbitally
degenerate ground state would  the $d_{xz}$, $d_{yz}$ doublet.  In
an ideal square pyramidal geometry, where the cation sits in the
center of the square plane of the pyramid, this doublet is indeed
the ground state.  However, for $V^{4+}$O$_5$ square pyramids, the
$V^{4+}$ cation is significantly displaced from this square plane
which lowers the energy of the $d_{xy}$ singlet relative to that
of the doublet. Consequently, the detailed structural arrangement
needs to be taken into account to determine the orbital ground
state. Ohama \emph{et. al} \cite{Ohama} calculated the crystal
field splitting as a function of displacement of the cation from
the basal plane under the assumption of a ideal square pyramid
with equal V-O distances of 1.9 \AA.  This calculation suggests
that when the distance exceeds about 0.35 \AA, the $d_{xy}$
singlet is the crystal field ground state.  As the true distance
for K$_2$V$_3$O$_8$ is 0.55~\AA, this calculation suggests a
non-degenerate singlet crystal field ground state ruling out the
above scenario of a mixed spin-orbital mode.

To rule out the possibility of low-lying crystal field modes, we
must consider the splitting between the $d_{xy}$ ground state and
the $d_{xz}$, $d_{yz}$ lowest excited states in a more
quantitative sense. Recent interest in S=1/2 systems where quantum
effects are significant have resulted in several materials
exhibiting V$^{4+}$O$_5$ square pyramidal local structure
\cite{Ueda}. Detailed molecular orbital calculations for
$\alpha^{\prime}$-NaV$_2$O$_5$, CaV$_2$O$_5$, and MgV$_2$O$_5$
\cite{Koo} all of which exhibit square pyramidal geometry similar
to K$_2$V$_3$O$_8$ confirms the $d_{xy}$ ground state and
estimates a rather large splitting to the lowest excited state
ranging from 600-900 meV.  Evidence of this large crystal field
splitting is also seen in band structure calculations for several
square-pyramidal V$^{4+}$ vanadates where the $d_{xy}$ bands are
well separated from the remainder of the $d$ bands
\cite{Pickett,Valenti1,Smolinskii,Valenti2,Rosner}. Similar
calculations performed on K$_2$V$_3$O$_8$ \cite{Rai} show a very
similar splitting of the $d_{xy}$ bands suggesting a rather large
crystal field splitting.  One particular compound of note is
Li$_2$VOSiO$_4$ \cite{Melzi} which has a crystal structure where
V-O distances within the square pyramid are almost identical to
K$_2$V$_3$O$_8$.  For this compound, an experimental estimate of
the splitting between the ground state and first crystal field
excited state yielded a splitting of $\sim$150 meV \cite{Melzi}
while LDA band structure calculations estimated this value to be
$\sim$400 meV \cite{Rosner}. The local structural similarity
between K$_2$V$_3$O$_8$ and these other vanadates together with
the evidence for rather large splitting in all of these materials,
is very strong evidence of splitting between the $d_{xy}$ ground
state and the $d_{xz}$,$d_{yz}$ lowest lying excited state for
K$_2$V$_3$O$_8$ which is on the order of hundreds of meV. A
splitting of this magnitude together with the non-degenerate
orbital ground state eliminates orbital effects from playing a
significant role on the mode splitting observed at much lower
energies ($\sim$3 meV).  As a caveat, it is important to note that
the crystal field splitting in K$_2$V$_3$O$_8$ has not been
directly measured experimentally and the above arguments are based
purely on comparison with other related vanadate systems together
with band structure calculations.

\subsection{Dilution/Randomness}

The measured anomalies near the zone boundary in the magnetic
excitation spectrum of K$_2$V$_3$O$_8$ are strikingly similar to
the calculated excitation spectrum for site diluted 2d
antiferromagnets \cite{disorder1,disorder2}.  Studies of the
effects of disorder on lattice models \cite{orbach_review} showed
the development of localized modes, named fractons
\cite{disorder3} with increasing disorder. Similar localized modes
were predicted to exist in studies of the effects of disorder on
Heisenberg ferromagnets and antiferromagnets \cite{orbach_review}.
Fractons have little impact on the excitation spectrum at longer
wavelengths as this effectively averages over enough sites to make
dilution effects negligible. However, as the characteristic
distance becomes smaller on approaching the zone boundary, the
influence of these fractons becomes much more significant and
theoretical predictions \cite{Aharony} suggest a multi-peaked
spectrum on approaching the zone boundary. Experimentally,
inelastic neutron scattering experiments on the 3d
antiferromagnet, Mn$_x$Zn$_{1-x}$F$_2$ \cite{Uemura} and the 2d
antiferromagnet, Rb$_2$Mn$_x$Mg$_{1-x}$F$_4$ \cite{Cowley},
provided clear evidence for the multi-peaked nature of
S(\textbf{Q},$\omega$) in the presence of substantial disorder
with concentrations in both cases very close to the percolation
limit.

Recently, detailed calculations of S(Q,$\omega$) have been
presented for the site diluted 2d Heisenberg antiferromagnet
\cite{disorder1,disorder2}.  These calculations show behavior
which is similar to the undiluted case at long wavelengths, albeit
with a renormalized spin-wave velocity \cite{disorder1}, with a
crossover on approaching the zone boundary to a multi-peaked
spectrum \cite{disorder1,disorder2} the form of which is very
similar to the experimental observations presented here for
K$_2$V$_3$O$_8$.  Specifically, at the zone boundary, the
theoretical predictions show two modes at weaker doping
concentrations, with a splitting between the modes which is
largely independent of doping level.  This theoretical splitting
is $\sim$0.6 meV using the experimental parameters of
K$_2$V$_3$O$_8$ which can be compared to the experimentally
measured splitting in K$_2$V$_3$O$_8$ of 0.45 meV.  These values
are in fairly good agreement, particularly given that anisotropy,
interplanar coupling, and quantum effects haven't been accounted
for in the theoretical models \cite{disorder1,disorder2}.  There
are, however, some differences between the experimental
observations and the theoretical predictions which should be
pointed out.  First, the calculated S(\textbf{Q},$\omega$)
\cite{disorder1,disorder2} contains two modes on approaching both
the ($\pi$/2,$\pi$/2) and ($\pi$,0) zone boundary points while the
experimental observations only show a clear splitting near
($\pi$/2,$\pi$/2) with what is best described as broadening near
($\pi$,0).  In addition, the calculations shown to date make no
predictions of how S(\textbf{Q},$\omega$) varies along the zone
boundary direction ($\pi$/2,$\pi$/2)-($\pi$,0). Consequently, it
is unclear as to whether the experimentally observed spectral
weight shift can be described by this model. Clearly, further
theoretical work is needed to determine if the qualitative
comparison to the site dilution model is able to describe the
anomalous features seen in the excitation spectrum of
K$_2$V$_3$O$_8$.

The obvious question that arises, however, is why there would be
significant disorder in K$_2$V$_3$O$_8$. This material is a mixed
valence material with an average valence of 4.667, containing both
magnetic V$^{4+}$ and non-magnetic V$^{5+}$. Crystallographically,
the three Vanadium positions in each formula unit are comprised of
2 tetrahedrally coordinated sites and 1 pyramidally coordinated
site.  The simplest manner of getting the correct overall valence
is to have the V$^{5+}$ located on the tetrahedral site and the
V$^{4+}$ on the pyramidal site and this is the assumed charge
ordering scheme. However, there is no strict experimental proof
that such an ordering occurs in K$_2$V$_3$O$_8$ and some valence
mixing on the pyramidal site could result in disorder effects.
Other mixed valence vanadates, most notably
$\alpha^{\prime}$-NaV$_2$O$_5$ are strongly influenced by charge
disproportionation and the resulting charge degrees of freedom
\cite{Smolinskii,nav2o5b}.  More detailed experimental studies are
necessary to explore the possibility of site disorder in the low
temperature state of K$_2$V$_3$O$_8$.

\subsection{Structural Distortion}

One final experimental note of potential importance is the recent
observation of a higher temperature phase transition (T$_c\sim$110
K) in K$_2$V$_3$O$_8$ \cite{Sales}.  This transition is clearly
structural in nature, as evidenced by changes in the vibrational
properties \cite{Choi}.
%Crystallographic studies of the low
%temperature state have yet to be competed making it difficult to
%estimate the impact of this transition on the low temperature
%magnetic excitation spectrum.
Compounds such as K$_2$V$_3$O$_8$ with the fresnoite structure are
typically sensitive to displacive structural phase transitions
resulting in incommensurately modulated structures \cite{Withers}.
In most systems, the incommensurate modulation is along the
(1,1,0) crystallographic direction and usually results in a
modulation wavevector of \textbf{Q}$\sim$0.3(1,1,0) or
\textbf{Q}$\sim$0.3(1,1,0)+0.5$c^{*}$. Very recent
crystallographic measurements on K$_2$V$_3$O$_8$ indicates the
presence of such a displacive transition with superlattice peaks
seen in x-ray diffraction studies with wavevector
\textbf{Q}$\sim$0.3(1,1,0)+0.5$c^{*}$ \cite{Chako}.
Experimentally, no direct evidence for this changed periodicity is
observed in the inelastic neutron scattering measurements.
%However, we do note that
%deviations from the expectations of spin wave theory do seem to
%occur at about 0.3\textbf{Q} along both (h,0,0), split mode begins
%to appear at about (1$\pm$0.3,0,0) (see Fig. \ref{finaldisp}), and
%along (h,h,0) where the mode broadening begins at about
%(0.3,1+0.3,0) (see Fig. \ref{psi_h1h0}). A larger unit cell could
%produce additional "optic" spin wave branches which may be another
%explanation for the observation of two modes near the
%antiferromagnetic zone boundary.
The presence of an incommensurate structural modulation will
affect the interactions between near-neighbor spins via a
long-wavelength modulation of the superexchange interactions. The
effects of periodic modulations of exchange interactions has
received some attention in recent years in investigating the
effects of stripe order on the magnetic interactions in cuprates
\cite{stripe1,stripe2,stripe3}. A recent theoretical investigation
described the modification of the Hamiltonian required to account
for such an exchange constants modulation in the limit of weak
modulation strength \cite{Zaliznyak}. This study suggests that
simple model systems, for instance, conventional antiferromagnets
such as K$_2$V$_3$O$_8$, would be insensitive to small
long-wavelength modulations of the exchange constants.  It should
be noted, however, that this study focussed on the ground state
properties and not more subtle changes to the spin-wave spectrum.
In addition, this study assumes a weak modulation of the exchange
which need not be the case. More theoretical effort is needed to
estimate the strength of the expected modulation for the
particular case of K$_2$V$_3$O$_8$ and the resulting effect of
such a modulation on the spin-wave spectrum.

Interestingly, the material Rb$_2$V$_3$O$_8$ which is
isostructural at room temperature, also exhibits a structural
phase transition with a very different modulation of
\textbf{Q}$\sim$0.16$c^{*}$ \cite{Withers}.  These distortions are
explained, within the rigid unit mode analysis, as resulting from
rotations of the V$^{5+}$O$_4$ tetrahedra such as to uniformly
raise or lower the V$^{4+}$O$_5$ pyramidal network along the
$c$-axis \cite{Withers}. The proposed modulation does not affect
the V$^{4+}$-O-O-V$^{4+}$ superexchange pathway within the 2d
planes only affecting the separation between layers and, hence,
the interplanar coupling. One would expect such a distortion to
have little effect on the excitation spectrum which is
predominately two-dimensional in character.  Comparison of the
magnetic excitations in Rb$_2$V$_3$O$_8$ to those of
K$_2$V$_3$O$_8$, where the modulation contains a large in-plane
component,  could shed some light on the nature of the observed
mode splitting.

\section{Conclusions}

In conclusion, we have performed detailed inelastic neutron
scattering studies of the magnetic excitation spectrum of the 2d
antiferromagnet, K$_2$V$_3$O$_8$ using a combination of
triple-axis and time-of-flight experiments.  The long-wavelength
region of the excitation spectrum confirms that K$_2$V$_3$O$_8$ is
an excellent example of a QSLHAF with a very small near-neighbor
coupling constant $J$=1.08$\pm$0.03 meV.  In addition, we were
able to confirm the presence of a very small anisotropy gap in the
excitation spectrum of 72$\pm$9 $\mu$eV.  Under the assumption of
the previously suggested Hamiltonian with the
Dzyaloshinskii-Moriya interaction set to
D$_z$/$\widetilde{J}$=0.04, we determine a c-axis exchange
anisotropy of E/$\widetilde{J}$=0.0012$\pm$0.0001 in very good
agreement with the values estimated based on the location of the
field-induced phase transitions \cite{Lumsden}.  Finally, the
dispersion along the c-axis was measured indicating a
ferromagnetic interplanar coupling J$_c$=-0.0036$\pm$0.0006 meV
(using J$_c$/$\widetilde{J}$=-0.0028$\pm$0.0005 and
$\widetilde{J}$=1.275$\pm$0.03) demonstrating that K$_2$V$_3$O$_8$
is a very good two-dimensional material.

As we approach the zone boundary, however, the previously
determined model no longer adequately describes the measured
excitation spectrum.  The excitations near the ($\pi$,0) zone
boundary point are broadened while near the ($\pi$/2,$\pi$/2) zone
boundary point we observe two clear modes rather evenly split
around the expected position of the single mode based on
measurements in long-wavelength limit and the predictions of
spin-wave theory. The upper of the two modes has a peculiar
Q-dependence in that it only has significant intensity near the
zone boundary. In addition, the intensity of the upper mode is
seen to fall off continuously as we move along the zone boundary
from ($\pi$/2,$\pi$/2) to ($\pi$,0). We have discussed several
possible explanations for the split mode near the zone boundary.
Among these explanations, the best agreement with the data is the
case of disorder on the 2d lattice although we have no other
evidence for disorder in this material. We hope that these
measurements will stimulate additional effort both experimentally
and theoretically to attempt to explain the peculiar zone boundary
properties of K$_2$V$_3$O$_8$.

\acknowledgements We would like to thank A. Zheludev and A.L.
Chernyshev for helpful discussions and to Mark Telling and Felix
Altofer for technical assistance. The triple-axis data was fit
using the Reslib software package, A. Zheludev and the
time-of-flight data was fit using routines derived from MSLICE, R.
Coldea.  We would like to thank A. Joshi, M. Ma, and F.C. Zhang
for calculating the crystal field energy levels. Research
sponsored by the Division of Materials Sciences and Engineering,
Office of Basic Energy Sciences, U.S. Department of Energy, under
contract DE-AC05-00OR22725 with Oak Ridge National Laboratory,
managed and operated by UT-Battelle, LLC. Work in London was
supported by a Wolfson Royal Society Research Merit Award.  D.A.T.
acknowledges the support by UK EPSRC Grant No. GR/N35038/01 for
IRIS measurement time.

\end{document}